\documentclass[%
 reprint,
 amsmath,amssymb,
 aps,
 prd,
]{revtex4-2}

\usepackage{graphicx}
\usepackage{color}
\usepackage{dcolumn}
\usepackage{bm}
\usepackage{hyperref}
\usepackage{subfigure}
\usepackage{acronym}
\usepackage{CJK}
\usepackage{orcidlink}

\newacro{GW}{gravitational wave}
\newacro{BBH}{binary black hole}
\newacro{MBHB}{massive black hole binary}
\newacroplural{MBHB}[MBHBs]{massive black hole binaries}
\newacro{GB}{Galactic binary}
\newacroplural{GB}[GBs]{Galactic binaries}
\newacro{SNR}{signal-to-noise ratio}
\newacro{PTA}{pulsar timing array}
\newacro{TDI}{time delay interferometry}
\newacro{PSD}{power spectral density}
\newacro{FAR}{false alarm rate}
\newacro{FAP}{false alarm probability}

\begin{document}
\begin{CJK*}{UTF8}{gbsn}

\preprint{APS/123-QED}

\title{Evaluating statistical significance for massive black hole binary mergers with space-based gravitational wave detectors}

\author{Hong-Yu Chen(陈洪昱)\orcidlink{0009-0006-2843-7409}$^1$}
\author{En-Kun Li(李恩坤)\orcidlink{0000-0002-3186-8721}$^1$}
\author{Yi-Ming Hu(胡一鸣)\orcidlink{0000-0002-7869-0174}$^{1,}$}\email{huyiming@sysu.edu.cn}
\affiliation{$^1$MOE Key Laboratory of TianQin Mission, %
            TianQin Research Center for Gravitational Physics $\&$ School of Physics and Astronomy, %
            Frontiers Science Center for TianQin, %
            Gravitational Wave Research Center of CNSA, %
            Sun Yat-sen University (Zhuhai Campus), %
            Zhuhai, 519082, China}
            
\date{\today}

\begin{abstract}
Important scientific discoveries should be backed by high statistical significance. 
In the 2030s, multiple space-based gravitational wave detectors are expected to operate.
While many works aim to achieve quick and reliable detection and parameter estimation of millihertz gravitational wave sources, dedicated studies are lacking to assess the significance of space-based detectors.
In this work, we propose a framework to assess the statistical significance of massive black hole binary (MBHB) detections with space-based gravitational wave detectors.
We apply this algorithm to simulated data with Gaussian stationary noise and the complex LDC-2a dataset to measure the false alarm rate and significance of MBHB signals.
We also analyze factors affecting the significance of MBHBs and design a method to mitigate multi-source confusion interference.
In Gaussian noise conditions, MBHBs with a signal-to-noise ratio of about 7 can achieve $3 \sigma$ significance, and those with a signal-to-noise ratio of about 8 achieve $4 \sigma$.
Our analysis demonstrates that all MBHB signals in the LDC-2a dataset have a significance exceeding $4.62 \sigma$.
\end{abstract}

\maketitle

\end{CJK*}

\section{\label{sec:Introduction}Introduction}

Scientific discoveries should be statistically confirmed.
When asserting a signal's existence, the null hypothesis probability (the likelihood of obtaining the data if no signal is present) must stay below a predetermined threshold. 
The discovery threshold is commonly set at $5\sigma$, which corresponds to the probability of a value being 5 or more standard deviations from the mean of a Gaussian distribution \cite{Lyons:2013yja}.

Recent years have seen significant advancements in the field of \ac{GW} astronomy, with nearly a hundred \ac{GW} events claimed by ground-based \ac{GW} detectors \cite{LIGOScientific:2018mvr,LIGOScientific:2020ibl,LIGOScientific:2021djp}.
Subsequently, \acp{PTA} have also contributed to the field by detecting the background of \acp{GW} \cite{NANOGrav:2023gor,EPTA:2023fyk,Reardon:2023gzh,Xu:2023wog}. 
In the 2030s, space-based \ac{GW} detectors are expected to further revolutionize this research area \cite{TianQin:2015yph,LISA:2017pwj,Hu:2017mde,Gong:2021gvw}. 
Operating in space and covering a lower frequency spectrum, these detectors will likely enhance our understanding of the cosmos.

One of the preliminary sources of space-based \ac{GW} detectors is \ac{MBHB} \cite{Wang:2019ryf,Katz:2018dgn}.
The space-based \ac{GW} detectors are expected to observe the \ac{MBHB} events with redshift up to $z\sim 20$. 
Detecting \acp{GW} from \ac{MBHB} mergers will be a significant breakthrough in astronomy and cosmology.
It will enable unprecedented tests of the theory of general relativity \cite{Berti:2004bd, Berti:2005qd, Gair:2012nm, Yagi:2016jml}, offer valuable insights into black hole formation and evolution \cite{Sesana:2007sh, Inayoshi:2019fun}, and provide important information about the universe's expansion \cite{Schutz:1986gp, Hughes:2001ya, Tamanini:2016zlh, Zhu:2021aat}. 
With so many important scientific applications deeply tied to the detection, the stakes are too high for us not to be extremely confident that the detections are genuine. 
Thus, providing a reliable assessment of the significance and \ac{FAR} of \ac{MBHB} signals is crucial for their application in astronomical and fundamental physics research.

Ground-based \ac{GW} detectors have developed various methods for evaluating significance, including the time-shift method \cite{Babak:2012zx} and the Poisson estimation method \cite{Cannon:2012zt}. 
However, no research team has yet explored significance evaluation methods for space-based \ac{GW} detectors. 
Space-based detectors' unique data characteristics complicate this process.
In the millihertz frequency band, \ac{GW} signals exhibit longer durations, causing multi-source confusion \cite{Bender:1997hs,Robson:2017ayy,Littenberg:2023xpl}. 
The satellite's orbit around its geometric center as well as the Sun creates a complex time-varying response, increasing the time and effort required to model the signals \cite{Vallisneri:2004bn,Marsat:2020rtl,Li:2023szq}. 
Meanwhile, for signals with generally high \ac{SNR}, more details like higher harmonics are no longer negligible, adding complexities \cite{Pitte:2023ltw}.
These factors complicate data analysis for space-based \ac{GW} detectors and necessitate the development of new methods.
Various methodologies have been developed to address the challenges of high-dimensional parameters and multi-source confusion in space-based \ac{GW} detection, such as wavelet denoising \cite{Cornish:2021smq}, block-Gibbs \cite{Deng:2025wgk}, novel maximum likelihood estimate \cite{Strub:2024kbe}, Reversible Jump Monte-Carlo Markov chains \cite{Littenberg:2020bxy,Littenberg:2023xpl,Katz:2024oqg}, and near real-time analysis \cite{Chen:2023qga}.
However, none of them have identified the signals they found or studied their \ac{FAR} and significance.

This work aims to develop a significance evaluation algorithm suitable for space-based \ac{GW} detectors, specifically for the detection of \ac{MBHB} mergers.
The estimate of significance is contingent on the search method, and we aim to push the limit with original data, with minimal assumptions on the statistical properties of the noise. 
In this study, we will select suitable time-shift intervals for space-based \ac{GW} detectors and construct a significance evaluation pipeline based on the iterative Nelder-Mead algorithm.

In addition, we will apply our algorithm to both Gaussian stationary noise and the more realistic LDC-2a dataset \footnote{\url{https://lisa-ldc.lal.in2p3.fr/challenge2a}}. The latter presents multi-source confusion from numerous \acp{GB} and \acp{MBHB}. 
We will analyze how foreground noise and multi-source confusion affect the significance of \acp{MBHB}. This analysis can offer insights for future advanced ground-based \ac{GW} detectors, which may also face this issue.

This paper is organized as follows:
Section~\ref{sec:back} gives an overview of the theoretical background of significance and its evaluation.
In Section~\ref{sec:result}, we analyze both the simulated Gaussian noise data and the LDC-2a Sangria dataset separately. 
Finally, Section~\ref{sec:summary} summarizes the entire work.
Appendix~\ref{sec:GWdetectors} provides a brief introduction to future space-based \ac{GW} detectors. 

\section{\label{sec:back} Theoretical background}

\subsection{\label{sec:significance} Statistical significance}

Significance quantifies the likelihood that an observed result is genuine rather than a random fluctuation. 
When the \ac{FAP} is below the probability corresponding to several standard deviations away from the mean of a Gaussian distribution, the significance of the signal is quantified using this $\sigma$ value, as shown in Table~\ref{tab:Gaussiansignificance}.

\begin{table}[htb]
    \renewcommand{\arraystretch}{1.2}
    \caption{The relationship between the FAP and significance.}
    \label{tab:Gaussiansignificance}
    \begin{tabular}{c|c}
        \hline
        {\rm Significance} & {\rm FAP} \\
        \hline
        $1\sigma$ & $1/6.30$ \\
        $2\sigma$ & $1/43.9$ \\
        $3\sigma$ & $1/741$ \\
        $4\sigma$ & $1/\left(3.16 \times 10^4 \right)$ \\
        $5\sigma$ & $1/\left(3.49 \times 10^6 \right)$\\
        \hline
    \end{tabular}
\end{table}

Achieving such high precision \ac{FAP} is challenging. 
The time-shift method effectively addresses this by offsetting the two detectors' timestamps to decorrelate the data and construct long-duration equivalent background data. 
For ground-based detector networks, using this method with several days of observational data from three detectors allows for constructing background data spanning longer than the universe's age \cite{Davies:2020tsx}. 
Then, \ac{FAR} and \ac{FAP} can be easily calculated by searching for coincident triggers in the equivalent background data:
\begin{align}
    {\rm FAR} &= \frac{N_{\rm FA}}{T_{\rm tot}}, \\
    {\rm FAP} &= \frac{T_{\rm obs}\times N_{\rm FA}}{T_{\rm tot}},
\end{align}
where $N_{\rm FA}$ indicates the number of coincident triggers in the equivalent background data, $T_{\rm tot}$ and $T_{\rm obs}$ are the total time of equivalent background data and the actual observation time.

\subsection{\label{sec:pipeline} Assessment of significance}

\begin{figure*}[ht!]
    \centering
    \includegraphics[width=0.9\textwidth]{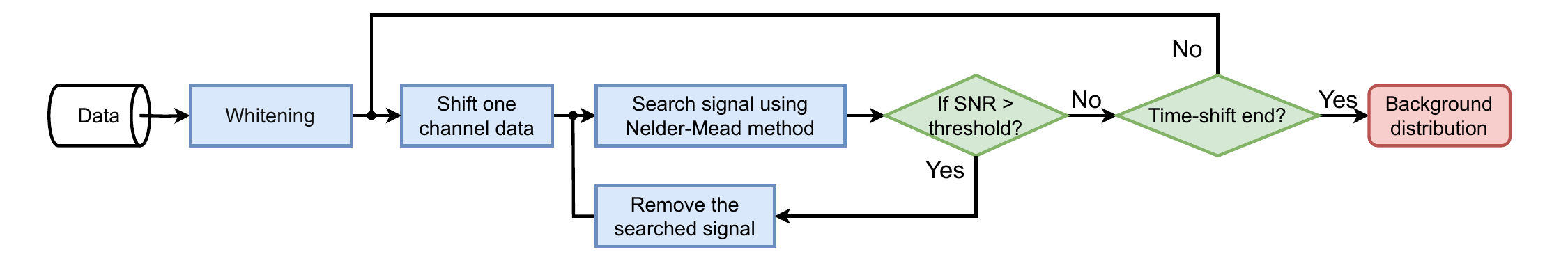}
    \caption{The significance evaluation pipeline for MBHBs in space-based GW detectors.}
    \label{fig:TimeshiftPipeline}
\end{figure*}

For space-based \ac{GW} detectors, we can adapt the time-shift concept with targeted optimizations, as depicted in~\ref{fig:TimeshiftPipeline}.
Unlike ground-based detectors that require multiple detectors for time-shifting, a triangular-shaped space-based \ac{GW} detector includes three interferometers.
These data streams can be rearranged into three noise-uncorrelated channels: A, E, and T \cite{Krolak:2004xp,Vallisneri:2004bn}.
Due to the T channel's low sensitivity, which barely affects the time-shift algorithm's results, we focus on $A$ and $E$. 
After whitening, we can construct equivalent background noise from the A and E channels. 
Considering the light-second scale of space-based detectors, as well as the timing errors, we've chosen a 1-minute interval for time-shifting. 
This choice is a compromise so that the two data channels are uncorrelated after the time-shift, meanwhile, one can construct a large enough background noise.
We remark that although for space-borne missions, glitches can be correlated and thus corrupt the sensitivity, they are less common and come from fewer sources than their ground-based counterparts \cite{McNamara:2008zz, LISAPathfinder:2019bxt, LISAPathfinder:2022awx}, so their effect are less severe. Thanks to the steady improvement in the glitch-cleaning pipeline \cite{Robson:2018jly, Baghi:2021tfd}, it can be expected that all major glitches can be identified and gated without harming channel independence.
Of course, this assumption was adopted for simplicity, and weaker glitches might be much harder to identify and could complicate the analysis.
Noise-monitoring channel can be used to flag the glitches \cite{Wu:2023rpn}, and, in the worst case, we can perform the time slide on channels from different detectors like TianQin and LISA.

Time-shift algorithms in ground-based \ac{GW} detectors contain two assumptions: inclusive background and exclusive background. The latter excludes all triggers with candidates that have a \ac{FAR} below a specified threshold, thus offering a more conservative description of the noise background distribution \cite{LIGOScientific:2016dsl, Capano:2017fia}. 
For space-based \ac{GW} detectors, MBHB signals, due to higher \acp{SNR} and longer durations, have a more significant impact on the background distribution. 
To eliminate this contamination and safely adopt a shorter time-shift interval, we confine our analysis to the exclusive background, i.e., the data set from which every identified \ac{MBHB} signal has been removed.

When identifying triggers in time-shifted data, we used the Nelder-Mead algorithm, a derivative-free multidimensional optimization method \cite{Nelder:1965zz, Lagarias:1998iqa}, which exceeds most search algorithms in speed.
It iteratively refines a simplex via four core operations, reflection, expansion, contraction, and shrinkage, to approach the optimal solution.
Instead of exploring the whole parameter space with a template bank, we use the Nelder-Mead method to iteratively refine an N-dimensional simplex, which converges to a maximum of the likelihood.
The method is quick since only a limited number of likelihood evaluations are required. 
Notice that it comes with a price that the method does not guarantee a global maximum and is dependent on the choice of the initial simplex \cite{Singer:2004anac}. 
In practice, a single \ac{GW} search with 500 iterations consumes about a few core minutes on a core for one year of data.
Its efficiency and ability to handle non-smooth or discontinuous functions without derivative information make it a suitable choice for fast optimization in practical applications.
To address the Nelder-Mead algorithm's sensitivity to initial values and susceptibility to local maxima, we optimize it via iterative loops, rendering it robust to initial value variations. Specifically, the standard algorithm is integrated into an iterative subtraction loop: post convergence, peaks with SNR exceeding the threshold are removed from the data; otherwise, the search restarts with random initial values. The process terminates upon multiple consecutive failures to detect new extrema, thus identifying all statistically significant peaks.

In the selection of the detection statistic, we chose an \ac{SNR} form based on the log-likelihood ratio. 
\begin{align}
    \rho &= \left\{\begin{matrix}
        \sqrt{2\times\ln\Lambda},  & \ln\Lambda>0\\
        0,  & \ln\Lambda\leq0
        \end{matrix}\right. \\
    \ln\Lambda &= \langle d \mid h \rangle - \frac{1}{2} \langle h \mid h \rangle,
\end{align}
where, $d$ is the data, and $h$ is the signal template.
This form of \ac{SNR} is better at suppressing triggers caused by noise compared to the optimal \ac{SNR} in data analysis \cite{Balasubramanian:1995bm}.
In this work, we utilized the IMRPhenomD template \cite{Husa:2015iqa,Khan:2015jqa,Digman:2022igm} and \ac{TDI} response \cite{Vallisneri:2004bn,Marsat:2020rtl,Li:2023szq} to generate \ac{MBHB} signals template.
In this work, we restrict the chirp mass range to $10^5\sim10^8 ~M_\odot$ and set an upper limit on the luminosity distance of 230 Gpc, which corresponds approximately to a redshift of 20.
This choice of upper limit is motivated by two factors: first, this range is around the edge of the horizon distance of space-borne detectors; secondly, astronomical models do not support mergers of MBHBs at larger redshifts.
Therefore, triggers from events with a larger redshift could only arise from false alarms, which usually should be associated with low SNR.
We believe that the choice of this upper limit in distance should not significantly alter our conclusions.
For the other parameters, we draw the starting points uniformly.

After each search, the corresponding signals will be subtracted from the data. The search is then repeated until no \ac{MBHB} triggers above the threshold are found.
By periodically time-shifting the channel data and conducting searches with the Nelder-Mead algorithm after each shift, we can obtain the background distribution of the false alarms.
In our algorithm, we allocate the analysis of time-shifted data across varying time intervals to different cores, enabling high parallelism in code execution.

\section{\label{sec:result}Result}

In the ideal case, where the signals can be exactly matched and removed, and when the noise is Gaussian and stationary, the statistical distribution of the false alarms can be analytically obtained.
In this section, we first test our algorithm in the Gaussian and stationary noise case to verify if the false alarm distribution of the Nelder-Mead search follows the Rayleigh distribution.
Then we apply the algorithm to the LDC-2a data set that contains multiple types of overlapping signals to test the robustness.

\subsection{\label{sec:GaussianNoise}Gaussian Noise}

\begin{figure*}[htbp]
    \centering
    \subfigure[Theoretical noise A/E channel PSD]{\includegraphics[width=0.45\textwidth]{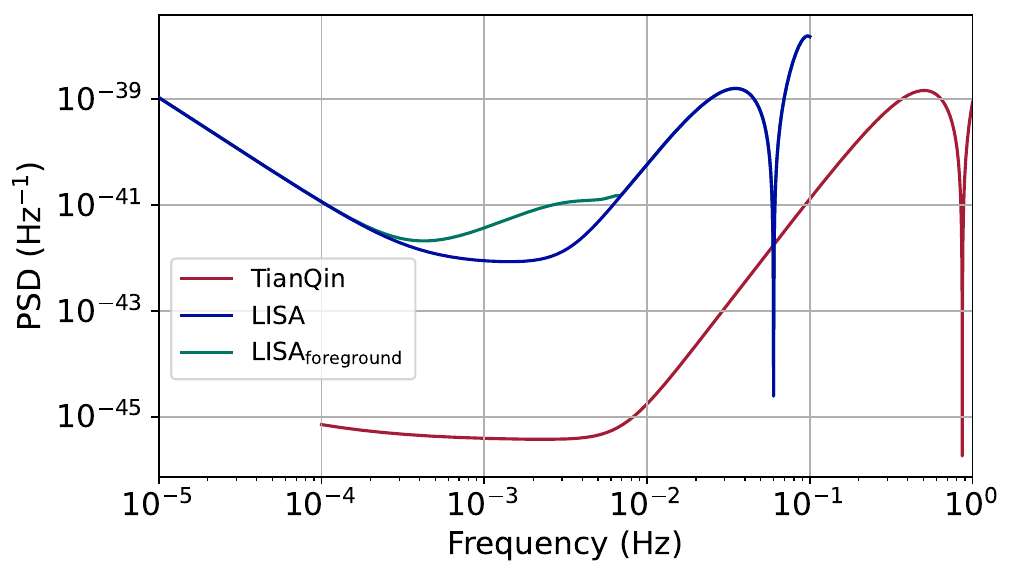}\label{fig:AchannelPSD}}
    \subfigure[SNR distribution of false alarms]{\includegraphics[width=0.45\textwidth]{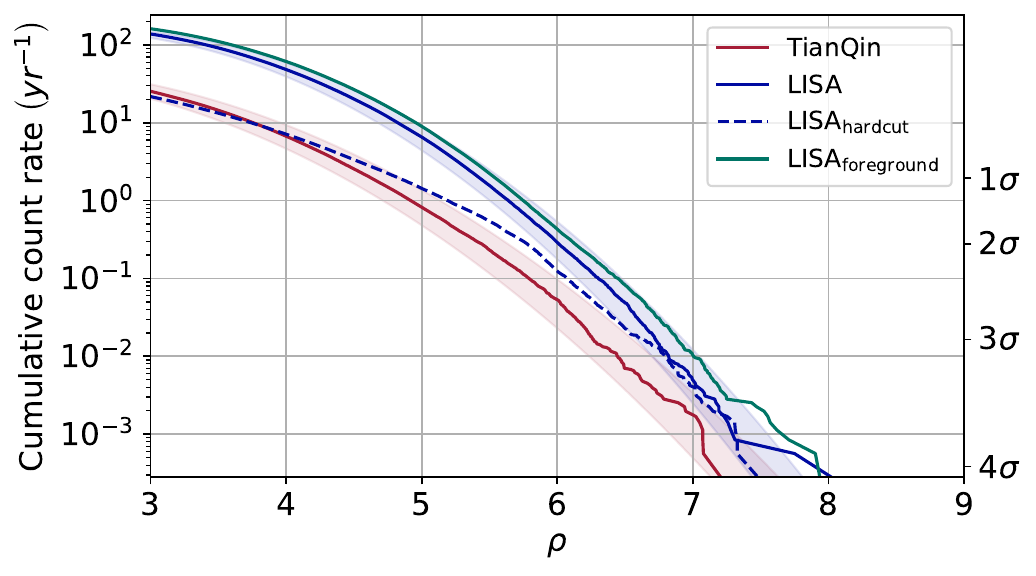}\label{fig:CCRofGaussianNoise}}
    \caption{The left panel shows the theoretical noise A/E channel PSD of TianQin (red), LISA (blue), and LISA with foreground (green). The right panel illustrates the SNR distribution of false alarms, which is compared with the survival function of the Rayleigh distribution (lighter areas). The blue dashed line denotes the results of LISA after the low-frequency hard cutoff. 
    The left-hand axis of the right panel shows the cumulative count rate (FAR) at different SNRs. The right-hand axis is the significance corresponding to different FARs.}
\end{figure*}

First, we consider the simplest scenario, where the data contains only Gaussian and stationary noise, representing the exclusive background for a single \ac{MBHB} source during observation. 
We generated 1 month of Gaussian instrumental noise for TianQin and LISA using the theoretical noise \ac{PSD} \cite{Prince:2002hp,Li:2023szq}, as shown in Fig.~\ref{fig:AchannelPSD}. 
Notably, due to LISA's superior low-frequency performance, it is more affected by the \ac{GB} foreground. Therefore, we also assessed the situation where LISA is impacted by the \ac{GB} foreground.
For simplicity, we consider the stationary \ac{GB} foreground here. In the next section, we will explore the more realistic impact of \ac{GB}.

By shifting 1-month observational data at 1-minute intervals, we moved the data 43,200 times to create equivalent background data spanning 3,600 years.
For each piece of data, by applying the Nelder-Mead algorithm, we can identify the trigger with the \ac{SNR} greater than 3.
All triggers with \ac{SNR} greater than 3 are accumulated and their distribution is shown in Fig.~\ref{fig:CCRofGaussianNoise}.
Since no authentic signals are expected in the time-shifted background, all the triggers can be treated as false alarms.
By setting an \ac{SNR} threshold, one can obtain the \ac{FAR} as the cumulative count rate of the false alarms which are associated with \ac{SNR} larger than the threshold.
In Fig.~\ref{fig:CCRofGaussianNoise}, we mark the cumulative count rate and significance for \ac{MBHB} signals with different \acp{SNR}.
Our results closely match the expected Rayleigh survival function, confirming the reliability of the proposed algorithm.
The lowest \ac{FAR} of $1/3600 ~{\rm yr}^{-1}$ corresponds to a significance level of $4.07 \sigma$.

Based on the \ac{SNR} distribution of false alarms from Gaussian noise, Table~\ref{tab:GaussianNoiseResult} presents the \ac{SNR} required for TianQin and LISA (both the instrumental noise itself and that includes foreground noise) to reach $3\sigma$ and $4\sigma$.
TianQin has a lower \ac{FAP} for \ac{MBHB} detection. 
After investigation, we conclude that such a difference stems from the difference in the low-frequency cutoff between the two missions. 
Compared with TianQin ($10^{-4}$Hz), LISA has a lower low-frequency cutoff for the frequency band ($10^{-5}$Hz). 
There are only a limited number of data points between the range of $10^{-5}-10^{-4}$Hz, and random fluctuations from Gaussian noise could fool the search algorithm into treating it as containing a low-SNR, high-mass MBHB merger.
The occurrence rate for lower-mass false alarms would be much smaller since it corresponds to higher frequencies and thus more data points are involved, making it harder for pure noise to mock the signal.
To verify this hypothesis, we applied a hard cutoff to LISA's low-frequency data at $10^{-4}$ Hz. It was found that the low-SNR false alarm rate of LISA decreased by an order of magnitude, approaching the results of TianQin.
This is clearly illustrated by the blue dashed line in Fig.~\ref{fig:CCRofGaussianNoise}. 
We can observe that after the hard cutoff, the false alarms at low SNRs of LISA are roughly comparable to those of TianQin, indicating that the excess of low-SNR false alarms of LISA primarily stems from Gaussian random fluctuations in the low-frequency data.
Meanwhile, in the high-SNR end, the dashed line approaches the solid line, indicating that the low-frequency data has a negligible contribution. 

\begin{table}[ht!]
    \renewcommand{\arraystretch}{1.2}
    \caption{The SNR required for MBHBs to reach a specified significance level under Gaussian noise for different detectors, assuming a 1-month observation.}
    \label{tab:GaussianNoiseResult}
    \begin{tabular}{c|c|c}
        \hline
        {\rm Detector} & $\rho_0^{3\sigma} $ & $\rho_0^{4\sigma} $ \\
        \hline
        TianQin & 6.26 & 7.15 \\
        LISA & 6.72 & 7.91 \\
        LISA$_{\rm foreground}$ & 6.88 & 7.92 \\
        \hline
    \end{tabular}
\end{table}

When analyzing Gaussian noise with an \ac{SNR} threshold of 3 for TianQin, the entire process required approximately 111 hours using 56 cores of an Intel Xeon Gold 6330 CPU.
For LISA cases with and without foregrounds, using the same CPU, the computational time was approximately 142 hours and 214 hours, respectively.

\subsection{\label{sec:LDC2a}LDC-2a Sangria Dataset}

We then consider a more realistic scenario, analyzing the significance of \ac{MBHB} signals within the LDC-2a Sangria dataset.
This dataset, spanning one year, includes Gaussian noise, 15 \ac{MBHB} signals, and approximately 30 million \ac{GB} signals.

In this analysis, we drop the assumption of the known \ac{PSD}.
So the first step would be estimating the \ac{PSD} from the Sangria dataset.
\citet{Cornish:2021smq} proposed methods to estimate two types of \ac{PSD} for the Sangria dataset: the Spectrum \ac{PSD} and the Smooth \ac{PSD}. 
The Spectrum \ac{PSD} treats \ac{GB} signals as non-Gaussian, non-stationary noise and includes their spectral lines in the \ac{PSD}, facilitating the removal of \ac{GB} interference during whitening. 
Conversely, the Smooth \ac{PSD} offers a \ac{PSD} without \ac{GB} spectral lines. We used both \acp{PSD} to whiten the Sangria dataset and compared the outcomes. 

To reduce contamination from detected signals, we obtained the \ac{MBHB} parameters in the Sangria dataset via low-latency detection from \citet{Cornish:2021smq}. 
SNR deviations ($\Delta \rho / \rho({\rm Smooth ~ PSD})$) of most signals under two PSD conditions are under 8\%, with the maximum deviation being 10.6\%.
After subtracting these signals from the original dataset, we analyzed the residuals. 
By conducting $525,600$ time-shifts, we generated hundreds of thousands of years of equivalent background data. The lowest \ac{FAR} of $1/525,600 ~{\rm yr}^{-1}$ corresponds to a significance level of $4.62\sigma$.

\begin{figure}[ht!]
    \centering
    \includegraphics[width=0.4\textwidth]{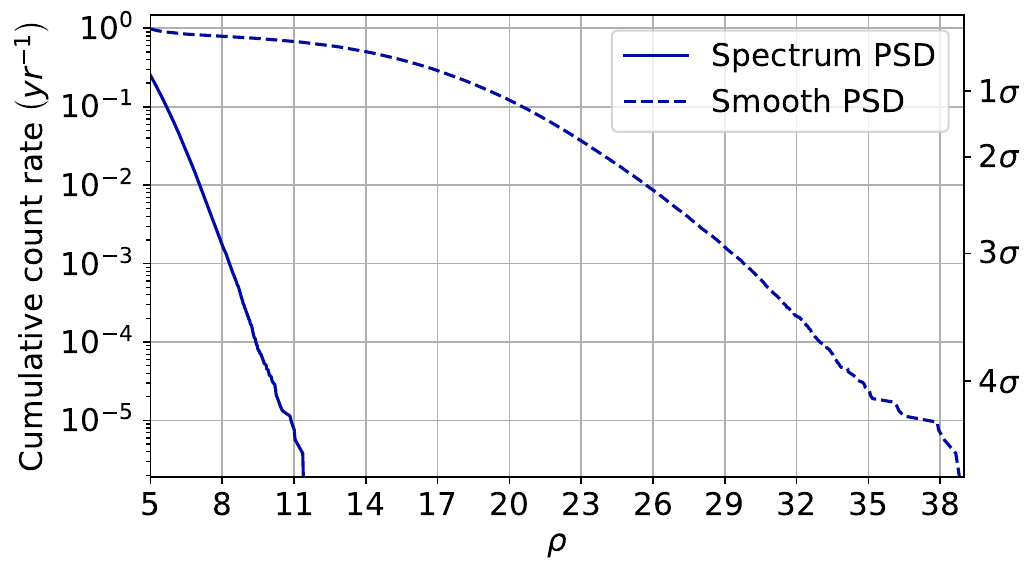}
    \caption{The SNR distribution of false alarms, similar with Fig.~\ref{fig:CCRofGaussianNoise}, but for Sangria dataset with two kinds of PSD.}
    \label{fig:CCRofLDC}
\end{figure}

As shown in Fig.~\ref{fig:CCRofLDC} and Table~\ref{tab:LDCResult}, the significance-to-SNR relationship differs greatly between the two \acp{PSD}. 
In the case of Spectrum \ac{PSD}, \ac{MBHB} achieves a $4\sigma$ significance at an \ac{SNR} of about 10, while for Smooth \ac{PSD}, it requires an \ac{SNR} near 35. 
This reveals that the Spectrum \ac{PSD} significantly suppresses false alarm \ac{SNR}, thus greatly boosting \ac{MBHB} significance. 
This method may apply to other multi-source confusion scenarios. 
Fortunately, as most \acp{MBHB} have large \ac{SNR}, after hundreds of thousands of time-shifts, no false alarm reaches the \ac{SNR} level of detected signals. Thus, we can confidently state that all 15 signals have a significance level exceeding $4.62 \sigma$.

\begin{figure*}[ht!]
    \centering
    \includegraphics[width=0.7\textwidth]{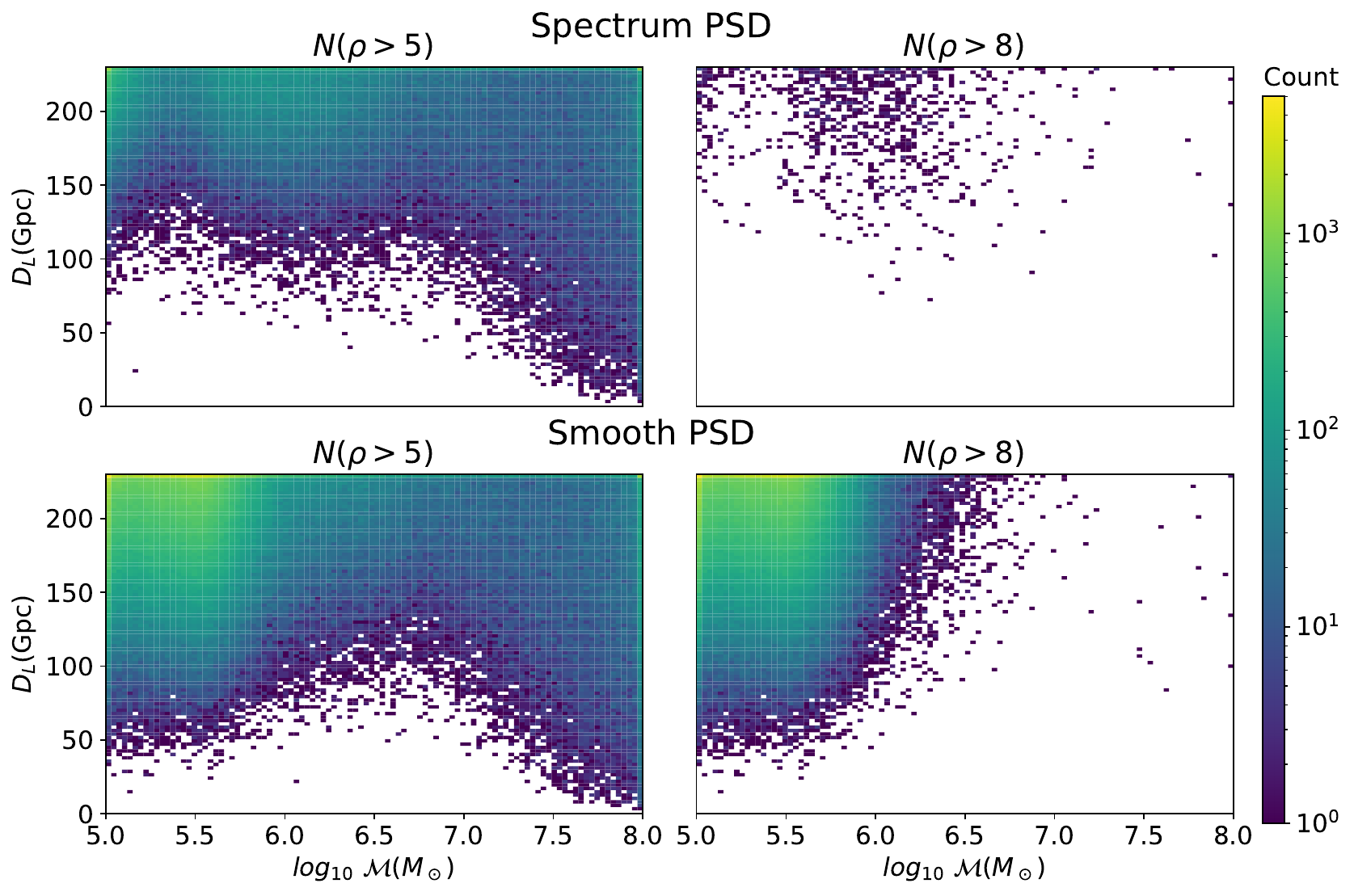}
    \caption{The histograms of the distribution of false alarms with $\rho \geq$ 5 (left panels) or 8 (right panels) in terms of chirp mass and luminosity distance for different PSDs.}
    \label{fig:DistributionofLDC}
\end{figure*}

To explore the causes of false alarms, we collected false alarms with $\rho \geq$ 5 or 8. 
These false alarms are mostly uniformly and randomly distributed in the parameter space but show patterns in chirp mass and luminosity distance. 
Fig.~\ref{fig:DistributionofLDC} shows that the \ac{SNR} of false alarms in the high-mass regime, primarily attributable to random fluctuations in the noise, is consistently below 8.
This conclusion is consistent with the results presented earlier (Fig.~\ref{fig:CCRofGaussianNoise}).
Moreover, the Spectrum \ac{PSD} effectively suppresses the impact of high \ac{SNR} \acp{GB} on \ac{MBHB} significance. In contrast, the Smooth \ac{PSD} results in significantly more false alarms with chirp masses of $10^5 \sim 10^6 ~M_\odot$.

Furthermore, Fig.~\ref{fig:DistributionofLDC} indicates that false alarms are not expected to be linked with nearby events ($\lesssim 50 {\rm Gpc}$).
Therefore, \ac{MBHB} candidates detected in this range are expected to be associated with higher significance, and it is possible to use this information to further boost their significance, so that even relatively weak signals could be unambiguously identified. 

For the LDC-2a dataset with an \ac{SNR} threshold of 5, under the Spectrum PSD condition, the analysis required about 198 hours using 336 cores across 6 Intel Xeon Gold 6330 CPUs, while the Smooth PSD condition consumed about 328 hours due to increased false alarms.
Our program is highly parallel and can utilize more cores to achieve reduced computation time.

\begin{table}[ht!]
    \renewcommand{\arraystretch}{1.2}
    \caption{The SNR required for MBHBs to reach a specified significance level, similar to Table~\ref{tab:GaussianNoiseResult}, but under the LDC-2a Sangria dataset with two kinds of PSDs.}
    \label{tab:LDCResult}
    \begin{tabular}{c|c|c}
        \hline
          & $\rho_0^{3\sigma} $ & $\rho_0^{4\sigma}$ \\
        \hline
        {\rm Spectrum PSD} & 8.15 & 10.09 \\
        {\rm Smooth PSD} & 29.32 & 34.64 \\
        \hline
    \end{tabular}
\end{table}

\section{\label{sec:summary}Summary}

Evaluating significance is essential to establish faith in future milli-Hertz \ac{GW} detections, while the estimation is tightly bound to the search algorithm adopted. 
Yet, studies on implementing significance estimation algorithms are still inadequate. 
In this work, we develop a fast, robust, and accurate significance evaluation algorithm for \ac{MBHB} mergers with space-based \ac{GW} detectors.
We leverage the time-shift method and the iterative Nelder-Mead algorithm, combined with the $A$, $E$, and $T$ channel characteristics of space-based detectors, to achieve significance evaluation for MBHBs under both ideal conditions and multi-source confusion scenarios.

We applied our significance evaluation pipeline to simulated Gaussian noise data and the LDC-2a dataset, and analyzed how factors like foreground noise and \ac{GB} signals affect the significance of \acp{MBHB}.
For the LDC-2a dataset with multi-source confusion, we found that treating high SNR \acp{GB} as spikes in the \ac{PSD} can greatly mitigate their contamination in false alarms, enhancing the significance of actual \acp{MBHB}.
Our analysis of one month of Gaussian and stationary data takes several thousand core hours, whereas analyzing a year of LDC-2a data requires tens of thousands of core hours. However, given the code's high parallelizability, the actual wall-clock time remains manageable in a multi-core parallel computing environment.

It should be noted that no glitches have been incorporated into the data of this work. In future work, we will conduct further investigations into the impacts of glitches on the FAR and potential mitigation strategies.
In addition, our current algorithm struggles to achieve $5\sigma$ significance unless given longer observation times or smaller time-shift intervals.
Enhancement is feasible, and we hereby propose several potential approaches to resolve this issue.
First, capitalizing on the unique properties of \ac{MBHB} signals, such as the \ac{SNR} accumulation patterns \cite{Chen:2024egt} and the mass-distance relationship, can further refine the filtering process. This approach eliminates potential false alarms and boosts the significance of candidates matching these features.
Second, the future multi-detector network will exponentially increase time-shift combinations through more observational channels, enabling a longer equivalent background data construction and a lower \ac{FAP} limit. 
Third, using a Poisson estimation method, we can extrapolate and estimate the significance of higher \ac{SNR} signals based on the existing significance-\ac{SNR} relationship.
Moreover, once we have a clear understanding of the noise properties, we can use this knowledge to model and simulate the noise for significance evaluation.

\section*{\label{sec:acknowledgment}Acknowledgment}
This work has been supported by the National Key Research and Development Program of China (No. 2023YFC2206700),   the Natural Science Foundation of China ( Grants  No.  12173104, No. 12261131504),  and the science research grants from the China Manned Space Project (CMS-CSST-2025-A13).
We would like to thank Jian-Dong Zhang, Han Wang, and Yu-Xin Yang for their helpful comments. 

\appendix
\section{\label{sec:GWdetectors}Detectors}

In the 2030s, there will be many space-based \ac{GW} detectors in operation, such as LISA\cite{LISA:2017pwj} from Europe, TianQin\cite{TianQin:2015yph} and Taiji\cite{Hu:2017mde} from China. These detectors all use an equilateral triangular configuration, emitting and receiving laser beams to each other, and searching \acp{GW} by detecting tiny changes in the distances between the satellites.

The TianQin detector is in a geocentric orbit and, due to its shorter arm length, is sensitive to signals in the frequency range of $10^{-4} \sim 1$ Hz. In contrast, LISA and Taiji are heliocentric orbit detectors, located $20^{\circ}$ ahead and behind Earth's orbit, and sensitive to signals in the frequency range of $10^{-5} \sim 10^{-1}$ Hz.

\bibliography{apssamp}

\end{document}